\documentclass[12pt]{article}

\usepackage{latexsym, amsmath, amssymb}
\usepackage{graphicx}
\usepackage{psfrag}
\usepackage[square, sort&compress, numbers, comma]{natbib}

\newcounter{subequation}[equation]
\makeatletter

\def\bcite{\@ifnextchar [{\@tempswatrue\@bcitex}{\@tempswafalse\@bcitex[]}}
\def\@bcitex[#1]#2{\if@filesw\immediate\write\@auxout{\string\citation{#2}}\fi
  \let\@bcitea\@empty
  \@bcite{\@for\@bciteb:=#2\do
    {\@bcitea\def\@bcitea{,\penalty\@m\ }%
     \def\@tempa##1##2\@nil{\edef\@bciteb{\if##1\space##2\else##1##2\fi}}%
     \expandafter\@tempa\@bciteb\@nil
     \@ifundefined{b@\@bciteb}{{\reset@font\bf ?}\@warning
       {Citation `\@bciteb' on page \thepage \space undefined}}%
     \hbox{\csname b@\@bciteb\endcsname}}}{#1}}
\def\@bcite#1#2{{#1\if@tempswa , #2\fi}}

\def\thesubequation{\theequation\@alph\c@subequation}
\def\@subeqnnum{{\rm (\thesubequation)}}
\def\slabel#1{\@bsphack\if@filesw {\let\thepage\relax
   \xdef\@gtempa{\write\@auxout{\string
      \newlabel{#1}{{\thesubequation}{\thepage}}}}}\@gtempa
   \if@nobreak \ifvmode\nobreak\fi\fi\fi\@esphack}
\def\subeqnarray{\stepcounter{equation}
\let\@currentlabel=\theequation\global\c@subequation\@ne
\global\@eqnswtrue
\global\@eqcnt\z@\tabskip\@centering\let\\=\@subeqncr
$$\halign to \displaywidth\bgroup\@eqnsel\hskip\@centering
  $\displaystyle\tabskip\z@{##}$&\global\@eqcnt\@ne
  \hskip 2\arraycolsep \hfil${##}$\hfil
  &\global\@eqcnt\tw@ \hskip 2\arraycolsep
  $\displaystyle\tabskip\z@{##}$\hfil
   \tabskip\@centering&\llap{##}\tabskip\z@\cr}
\def\endsubeqnarray{\@@subeqncr\egroup
                     $$\global\@ignoretrue}
\def\@subeqncr{{\ifnum0=`}\fi\@ifstar{\global\@eqpen\@M
    \@ysubeqncr}{\global\@eqpen\interdisplaylinepenalty \@ysubeqncr}}
\def\@ysubeqncr{\@ifnextchar [{\@xsubeqncr}{\@xsubeqncr[\z@]}}
\def\@xsubeqncr[#1]{\ifnum0=`{\fi}\@@subeqncr
   \noalign{\penalty\@eqpen\vskip\jot\vskip #1\relax}}
\def\@@subeqncr{\let\@tempa\relax
    \ifcase\@eqcnt \def\@tempa{& & &}\or \def\@tempa{& &}
      \else \def\@tempa{&}\fi
     \@tempa \if@eqnsw\@subeqnnum\refstepcounter{subequation}\fi
     \global\@eqnswtrue\global\@eqcnt\z@\cr}
\let\@ssubeqncr=\@subeqncr
\@namedef{subeqnarray*}{\def\@subeqncr{\nonumber\@ssubeqncr}\subeqnarray}
\@namedef{endsubeqnarray*}{\global\advance\c@equation\m@ne%
                           \nonumber\endsubeqnarray}

\renewcommand\maketitle{\par
  \begingroup
    \if@twocolumn
      \ifnum \col@number=\@ne
        \@maketitle
      \else
        \twocolumn[\@maketitle]%
      \fi
    \else
      \newpage
      \global\@topnum\z@   % Prevents figures from going at top of page.
      \@maketitle
    \fi
    \thispagestyle{plain}\@thanks
  \endgroup
  \setcounter{footnote}{0}%
  \global\let\thanks\relax
  \global\let\maketitle\relax
  \global\let\@maketitle\relax
  \global\let\@thanks\@empty
  \global\let\@author\@empty
  \global\let\@date\@empty
  \global\let\@title\@empty
  \global\let\title\relax
  \global\let\author\relax
  \global\let\date\relax
  \global\let\and\relax
}

\makeatother

\DeclareFontFamily{OT1}{rsfs11}{}
\DeclareFontShape{OT1}{rsfs11}{m}{n}{ <-> rsfs11 }{}
\DeclareMathAlphabet{\mathscript}{OT1}{rsfs11}{m}{n}

\numberwithin{equation}{section}

\newcommand{\gtlt}{\mathrel{\raise2.5pt\hbox{\oalign{$\scriptstyle>$\crcr
$\scriptstyle<$}}}}

\newcommand{\e}{{\mathrm e}}
\newcommand{\eg}{{\it e.g.}~}
\newcommand{\half}{\frac{1}{2}}
\newcommand{\ie}{{\it i.e.}~}

\newcommand{\bo}{\raise-0.4mm\hbox{$\Box$}}              % D'Alembertian

\newcommand{\pt}{\partial}

\newcommand{\be}{\begin{equation}}
\newcommand{\ee}{\end{equation}}
\renewcommand{\[}{\begin{equation}}
\renewcommand{\]}{\end{equation}}
\renewcommand{\(}{\left(}
\renewcommand{\)}{\right)}
\newcommand{\nn}{\nonumber}
\newcommand{\bea}{\begin{eqnarray}}
\newcommand{\eea}{\end{eqnarray}}
\newcommand{\bsea}{\begin{subeqnarray}}
\newcommand{\esea}{\end{subeqnarray}}

\renewcommand{\tt}{\rightarrow} % i.e.  \tt = 'tends to'
\newcommand{\sech}{\mathrm{sech}}

\newcommand{\rb}{\bar{r}}
\newcommand{\cF}{\mathcal{F}}
\newcommand{\cG}{\mathcal{G}}

\def\a{\alpha}
\def\b{\beta}

\renewcommand{\d}{\mathrm{d}}

\def\e{\epsilon}

\def\f{\phi}

\def\m{\mu}
\def\n{\nu}
\def\o{\omega}

\def\t{\tau}

\def\w{\omega}

\def\L{\Lambda}

\def\cF{{\cal F}}

\begin{document}

\begin{titlepage}
\begin{flushright}
%Imperial/TP/ \\
%KUL-TF-03/15\\
{\small
DAMTP-2006-114 \\[-0.5ex]
ITFA-2006-47 }
\end{flushright}
\vspace{.5cm}
\begin{center}
\baselineskip=16pt {\huge   Colliding Branes in Heterotic M-Theory
\\ }
\vspace{10mm}%27.mm
{\large Jean-Luc Lehners$^{\dag}$, Paul McFadden$^{\ddag}$ and
Neil Turok$^{\dag}$}
\vspace{15mm}%\vskip 7mm%1cm

{\small\it $^\dag$ DAMTP, CMS, Wilberforce Road, CB3 0WA,
Cambridge, UK.
\\ \vspace{6pt}
$^\ddag$ ITFA, Valckenierstraat 65, 1018XE Amsterdam, the Netherlands.} \\
\vspace{6mm}
%{\small
%$23^{\mathrm{rd}}$ November 2006
%}
\end{center}

\vspace{2mm} \abstract{\vspace{1mm} We study the collision of two
flat, parallel end-of-the-world branes in heterotic M-theory. By
insisting that there is no divergence in the Riemann curvature as
the collision approaches, we are able to single out a unique
solution possessing the local geometry of (2d compactified
Milne)/$\mathbb{Z}_2\times \mathbb{R}_3$, times a finite-volume
Calabi-Yau manifold in the vicinity of the collision. At a finite
time before and after the collision, a second type of singularity
appears momentarily on the negative-tension brane, representing
its bouncing off a zero of the bulk warp factor. We find this
singularity to be remarkably mild and easily regularised. The
various different cosmological solutions to heterotic M-theory
previously found by  other authors are shown to merely represent
different portions of a unique flat cosmological solution to
heterotic M-theory. }

\vspace{2mm} \vfill \hrule width 2.3cm \vspace{2mm}{\footnotesize
\noindent \hspace{-9mm}
 E-mail: \texttt{j.lehners@damtp.cam.ac.uk,
mcfadden@science.uva.nl, n.g.turok@damtp.cam.ac.uk.} }
\end{titlepage}

%\tableofcontents{}
%\newpage
\setcounter{page}{2}

\section{Introduction}

The riddle of the initial singularity is one of the most basic
challenges in cosmology. In standard four-dimensional general
relativity, the Riemann curvature diverges at the big bang
signalling an irretrievable breakdown of the theory. In
higher-dimensional string and M-theory, however, the nature of the
initial singularity is significantly altered.  Within the
higher-dimensional picture, the Riemann curvature may remain
bounded all the way to the singularity. In this situation, string
and M-theory corrections to the background geometry remain small,
allowing one to attempt to study the propagation of strings and
branes right up to, and perhaps even across, the singularity.

In this paper, we wish to study such a model of the cosmological
background spacetime, within an especially well-motivated
theoretical framework. Heterotic M-theory is built on the profound
correspondence between supergravity and strongly-coupled heterotic
string theory \cite{HW1, HW2}, and it remains one of the most
promising approaches to the unification of particle physics and
gravitation.  From the perspective of particle phenomenology,
heterotic M-theory makes full use of branes and of the extra
M-theory dimension, in order to solve the fundamental puzzles of
chirality and of the difference between the GUT and Planck scales.
From a cosmological perspective, heterotic M-theory provides a
setup in which standard model matter is localised on branes,
allowing the background density of such matter to remain finite at
the initial singularity.

We consider the collision of two flat, parallel end-of-the-world
branes within heterotic M-theory \cite{Khoury:2001wf, Seiberg}.
From the standpoint of the four-dimensional effective theory, this
event is a cosmological singularity of the usual catastrophic
type. Yet from the higher-dimensional standpoint, the situation is
far less singular. The density of standard model matter remains
finite. Moreover, even the bulk spacetime between the branes is
relatively well-behaved: as we shall show, the Riemann curvature
remains finite for all times away from the collision event itself.
In the case of trivial, toroidal compactifications of M-theory,
this phenomenon is well known. Here, the colliding-brane spacetime
is locally flat, a product of two-dimensional compactified Milne
space-time, modded out by $\mathbb{Z}_2$, with nine-dimensional
flat space. At the collision itself, the spacetime is
non-Hausdorff since one dimension momentarily disappears.
Nevertheless, the low-energy degrees of freedom, described near
the collision by winding $M2$ branes (or, equivalently, weakly
coupled heterotic strings), possess a regular evolution across the
singularity \cite{TPS}.

In more realistic models, six spatial dimensions are compactified
on a Calabi-Yau manifold. The dynamics are then described by a
five-dimensional effective field theory known as heterotic
M-theory, which is a consistent truncation of eleven-dimensional
supergravity coupled to the boundary branes \cite{LOSW1, LOSW2}.
Recently, an improved formulation has been developed which avoids
problematic terms involving squares of delta functions
\cite{Moss1, Moss2}, giving one greater confidence that classical
solutions of the five-dimensional effective theory do indeed
provide consistent M-theory backgrounds. In this paper, we shall
show that these equations possess a unique global solution
representing the collision of two flat end-of-the-world branes
which, in the vicinity of the collision, reduces to (2d
compactified Milne)/$\mathbb{Z}_2\times  \mathbb{R}_3$, times a
finite-volume Calabi-Yau manifold. In this solution, the Riemann
curvature is again bounded at all times away from the collision
event itself. Our solution offers the intriguing prospect of
modelling the big bang as a brane collision in a setup with a high
degree of physical realism.

The idea that the big bang was a brane collision in heterotic
M-theory was first proposed in \cite{Khoury:2001wf, Seiberg}.
However, the solution representing the approach and collision of
two branes has not so far been given in complete form. Important
steps towards such a solution were taken by Chamblin and Reall
\cite{Chamblin:1999ya}, who found a static solution for the bulk
geometry in which moving, spatially flat branes can be
consistently embedded. The bulk geometry possesses a timelike
virtual naked singularity lying `beyond' the negative-tension
boundary brane. Two qualitatively different solutions then exist
according to whether the boundary branes move through the static
bulk in the same, or in opposing, directions. Both of these
solutions, together with their time-reverses, are naturally
combined within our global solution. In a more recent paper
\cite{Gary}, Chen {\it et al.~}considered the solution where both
the positive- and negative-tension boundary branes move towards
the virtual singularity. They found an exact solution in a
convenient alternative coordinate system which is comoving with
the branes. However, Chen {\it et al.~}found that both branes
encounter the naked singularity, resulting in the annihilation of
the entire spacetime.

\begin{figure}[p]
\begin{center}
\hspace{-1cm}
\includegraphics[width=17cm]{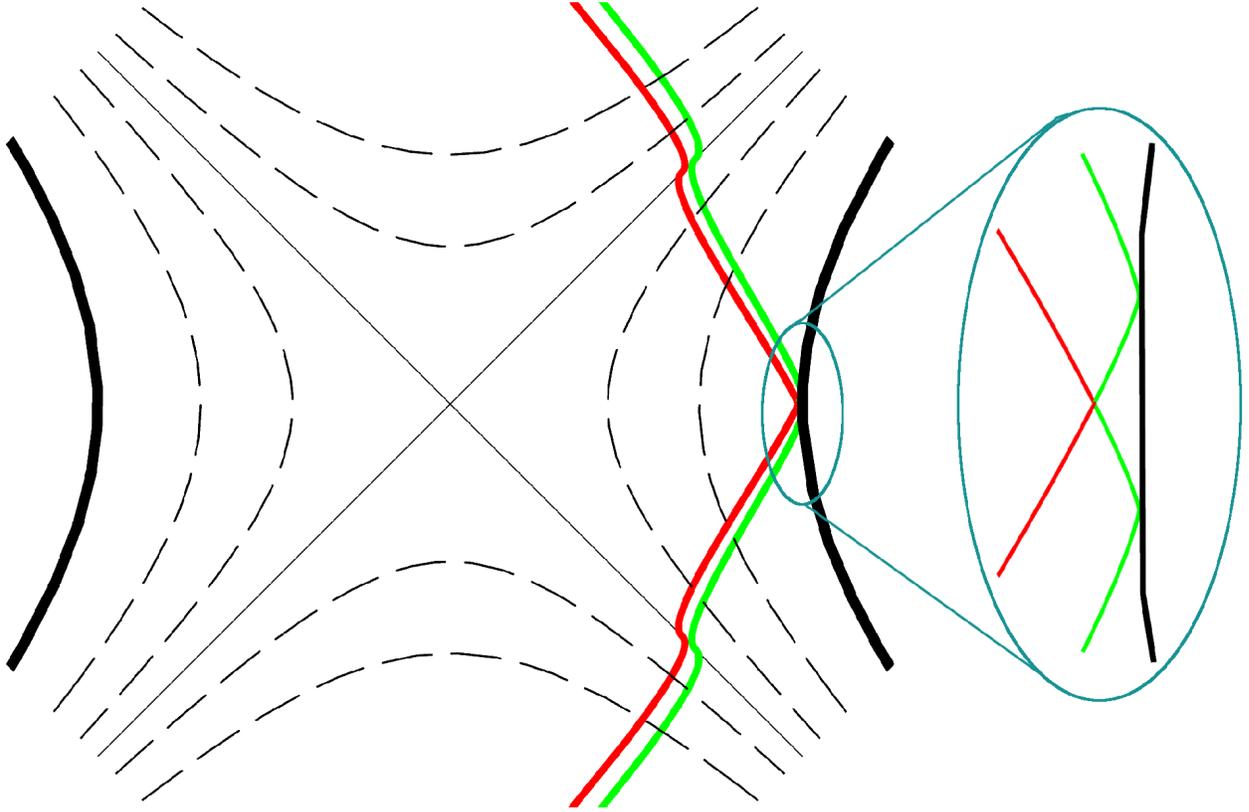}
\vspace{0.5cm} \caption{ \label{Kruskal} {\small
  A Kruskal plot of the entire solution, in the spirit of Chen {\it et al.~}\cite{Gary}.
The exact trajectories of the positive- and negative-tension
branes are plotted in red and green respectively, while the naked
singularity is indicated by a thick black line. Representative
orbits of the bulk Killing vector field are shown with dashed
lines, while the solid straight lines indicate the Boyer axes. The
bounces of the negative-tension brane off the naked singularity,
as well as the collision of the branes themselves, are shown at a
magnified scale in the inset. In this plot, we have chosen the
relative rapidity of the brane collision to be $2y_0 = 1$.
Analogous plots for greater collision rapidities may be found in
Figure \ref{doubleK} of Appendix C. } }
\end{center}
\end{figure}

In this paper, we shall argue for a different fate of the Chen
{\it et al.~}solution. We shall show that the naked singularity is
`repulsive' as far as the negative-tension brane is concerned, so
the brane approaches it with vanishing speed. The ensuing
singularity is extremely mild and easily regulated, for example by
introducing an arbitrarily small amount of matter on  %%V2
the negative-tension brane. The result is a smooth bounce, at
finite Calabi-Yau volume, so that the negative-tension brane
recoils from the singularity and collides with the
positive-tension brane. The end-of-the-world brane collision is
locally (2d compactified Milne)/$\mathbb{Z}_2\times 
\mathbb{R}_3$, with a finite-volume Calabi-Yau manifold.  Assuming
that M-theory can deal with this second singularity in the manner
described by \cite{TPS}, we can follow the system through. The
M-theory dimension then re-appears and the negative-tension brane
is thrown back towards the naked singularity. A second bounce of
the negative-tension brane occurs before the system continues into
the future with both branes expanding. The complete solution is
illustrated in the Kruskal diagram shown in Figure \nolinebreak \ref{Kruskal}.

We will present the single, global solution in two different
coordinate systems, each of which has some merit. In the first
system, which comoves with the branes so that they are kept at
fixed coordinate locations, only the bulk is dynamic. Here, we are
able to derive the solution as a series expansion in the relative
rapidity of the branes at the collision. This method has been
previously applied to solve for the bulk geometry and cosmological
perturbations of the Randall-Sundrum model \cite{McFadden}. The
chief virtue of this brane-comoving coordinate system is that it
simplifies the junction conditions on the branes, and allows for a
manageable treatment of cosmological perturbations.

In the second coordinate system, the branes move whereas the bulk
is static. The boundary conditions that we impose at the moment of
the collision, in conjunction with the assumed cosmological
symmetry and spatial flatness of the branes,
 allows us to derive a modified Birkhoff theorem singling
out a unique solution.  Away from the brane collision, this
solution has finite Riemann curvature throughout, at least after
suitable regularisation of the bounce of the negative-tension
brane. The ensuing evolution, before and after the latter event,
is described by Chamblin and Reall's two solutions
\cite{Chamblin:1999ya}, the second of which Chen {\it et
al.~}succeeded in re-writing in a brane-comoving coordinate
system.

The outline of this paper is as follows. In Section 2, we review
the standard static solution of heterotic M-theory and fix
notation. In Section 3, we derive the colliding brane solution in
brane-comoving coordinates, as an expansion in the rapidity of the
brane collision. In Section 4, we discuss the same solution in a
coordinate system in which the bulk is static and the branes are
moving.  Considering the relevant modified Friedmann equation, we
then show how the negative-tension brane bounces when an
arbitrarily small amount of matter is present on the brane. Three
appendices are devoted to technical matters. The first proves a
Birkhoff-like theorem for the bulk. The second derives the
equations governing the motion of branes in the bulk, and the
third discusses the Kruskal extension of the bulk geometry. These
calculations are used to make accurate plots of the brane
trajectories, like that shown in Figure \ref{Kruskal}.

\section{Heterotic M-theory}

Eleven-dimensional supergravity can be compactified on a
Calabi-Yau 3-fold to give a minimal five-dimensional supergravity
theory \cite{HW1, HW2, LOSW1, LOSW2}. Although the dimensional
reduction of the graviton and the 4-form flux generates a large
number of fields, it is consistent to retain only the
five-dimensional graviton, and a scalar $\f$ parameterising the
volume of the Calabi-Yau manifold (namely $V_{\mathrm{CY}} =
e^{\f}$). This theory can be consistently coupled to two
four-dimensional boundary brane actions, provided that one keeps,
in addition to gravity and the scalar, the components of the
4-form flux with all indices pointing in the Calabi-Yau
directions. This flux thus appears as a scalar in five dimensions
(albeit a constant one, as can be seen from its Bianchi identity
\cite{LOSW1, LOSW2}), and it leads to a potential for $\f.$ The
resulting dimensionally reduced effective action is given by \bea
S &=& \int_{5d} \sqrt{-g} \, [R - \frac{1}{2}(\pt \phi)^2 - 6\a^2
e^{-2 \phi}] \nn \\ [1ex] && + 12 \a \int_{4d, \, y=-y_0}
\sqrt{-g^-}\, e^{- \phi} - 12 \a \int_{4d, \, y=+y_0} \sqrt{-g^+}\,
e^{- \phi},
\label{Action5d} %\\ \nn
\eea where $\a$ is related to the number of units of 4-form
flux\footnote{Compared to \cite{LOSW1, LOSW2}, we have rescaled
$\a$ such that $\a =( \a_{\mathrm{LOSW}}/3\sqrt{2}).$}, and where
we have placed branes of opposite tensions at $y=\pm y_0$ ($y$
being the coordinate transverse to the branes). In brane-comoving
coordinates, the resulting equations of motion are \bea
\label{eom1} G_{ab} &=& \frac{1}{2}\f_{,a}\f_{,b} -
\frac{1}{4}g_{ab}\f^{,c}\f_{,c}-3 g_{ab}\,\a^2 e^{-2 \f} \nn \\
&& + 6 \a [\delta(y+y_0)-\delta(y-y_0)]\delta_a^{\m}\delta_b^{\n}g_{\m \n} \frac{1}{\sqrt{g_{yy}}}\,e^{-\f}, \\
\label{eom2} \Box \f &=& - 12\a^2 e^{-2 \f} + 12 \a
[\delta(y+y_0)-\delta(y-y_0)]\frac{1}{\sqrt{g_{yy}}}\,e^{-\f},
\eea where Latin indices run over all five dimensions, but Greek
indices run over only the four dimensions  to which the  $y$
coordinate is normal. The static domain wall vacuum solution is
given by \bea
\d s^2 &=& h^{2/5}(y)(\eta_{\m \n} \d x^{\m} \d x^{\n} + \d y^2),  \nn \\
e^{\f} &=& h^{6/5}(y), \label{domainwall} \nn \\
h(y) &=& 5\a(y+y_0)+c, \qquad -y_0 \leq y \leq y_0. \eea The $y$
coordinate is taken to span the orbifold $S^1/\mathbb{Z}_2$, with
fixed points at $\pm y_0$. In an `extended' picture of the
solution, obtained by $\mathbb{Z}_2$-reflecting  the solution
across the branes, there is a downward-pointing kink at $y=-y_0$
and an upward-pointing kink at $y=+y_0$. These ensure that the
appropriate Israel matching conditions are satisfied, with the
negative-tension brane being located at  $y=-y_0$ and the
positive-tension brane at  $y=+y_0$. The integration constant $c$
is arbitrary.

\section{A cosmological solution with colliding branes}

We now wish to generalise the static solution above to allow for
time dependence. In particular, we are interested in cosmological
solutions in which the two boundary branes collide, with the
five-dimensional spacetime geometry about the collision reducing
to (2d compactified Milne)$/ \mathbb{Z}_2 \times \mathbb{R}^3$. We
will therefore take as our metric ansatz \be \label{theusual} \d
s^2 = n^2(t,y)(-\d t^2 + t^2 \d y^2) + b^2(t,y)\, \d\vec{x}\,^2,
\ee where the $x^i$ span the three-dimensional spatial worldvolume
of the branes, which we are assuming to be flat.  This ansatz is
in fact the most general form consistent with three-dimensional
spatial homogeneity and isotropy, after we have made use of our
freedom to write the ($t$, $y$) part of the metric in conformally
flat Milne form. In the above, and throughout this section, we
will take the branes to be fixed at the constant coordinate
locations $y = \pm y_0$.

The functions $n$ and $b$, as well as the scalar field $\f$, are
arbitrary functions\footnote{Note in particular that we are not
assuming a factorised metric ansatz (\eg $n(t,y) = n_1(t)
n_2(y)$), since this generically leads to brane collisions in
which the Calabi-Yau manifold shrinks to zero size at the moment
of collision \cite{LS}.} of $t$ and $y$.  In order that the
collision be the `least singular' possible, we require that the
brane scale factors, as well as the Calabi-Yau volume, be finite
and non-zero at the collision. By an appropriate choice of units,
we will therefore demand that $n$, $b$ and $e^\phi$ all tend to
unity as $t \tt 0$.  In this way the geometry about the collision
reduces to the desired (2d compactified Milne)$/\mathbb{Z}_2
\times \mathbb{R}^3$ times a finite-volume Calabi-Yau manifold.
Since the five-dimensional part of this geometry is flat, our
solution will be protected from higher-derivative string or
M-theory corrections in the vicinity of the collision.

To solve the relevant equations of motion, (\ref{eom1}) and
(\ref{eom2}), we will perform a perturbative expansion in the
rapidity of the brane collision. This latter quantity is given by
$2y_0$ (see \eg \cite{Tolley:2003nx}), and so we are interested in
the case in which $y_0 \ll 1$, {\it i.e.}, the case in which the
relative velocity of the branes at the collision is small.  (See
\cite{McFadden} for an analogous procedure in the case of the
Randall-Sundrum model).

To implement this expansion, we first introduce the re-scaled time
and orbifold coordinates \be \omega = y/ y_0, \qquad T= y_0 t, \ee
so that the branes are now located at $\o = \pm 1$, independent of
the rapidity of the brane collision. For convenience, we will also
adopt the convention that primes denote derivatives with respect
to $\omega$ and dots denote derivatives with respect to $\ln{T}$,
{\it i.e.}, \be ' \equiv \pt_\o, \qquad \dot{\,} \equiv T \pt_T.
\ee In order to write down the junction conditions and the
Einstein equations, it is useful to work with the variables
$e^{\b} = b^3$ and $e^{\n} = y_0 n t = n T$.  The junction
conditions \cite{Israel}, valid when evaluated at $\omega= -1^+$
or $\omega = +1^-$, are then \be \nu' = \a\, e^{\nu-\phi}, \qquad
\beta' = 3 \a\,
e^{\nu-\phi}, \qquad \phi'= 6 \a\, e^{\nu-\phi}. \label{jct} \ee %%V2
Evaluating the $\phi$, $G^0_0+G^5_5$, $G^5_5$,
$G^0_0+G^5_5-(1/2)G^i_i$, and $G_{05}$ equations in the bulk, we
obtain the following set of equations: \bea \label{Mphi}
\phi''+\beta'\phi'+12 \a^2 e^{2\nu-2\phi} &=& y_0^2\,(\ddot{\phi}+\dot{\beta}\dot{\phi}),  \\
\label{MG00p55}
\beta''+\beta'^2+6\a^2 e^{2\nu-2\phi} &=& y_0^2\, (\ddot{\beta}+\dot{\beta}^2), \\
\label{MG55}
\frac{1}{3}\,\beta'^2+\beta'\nu'-\frac{1}{4}\,\phi'^2+3\a^2
e^{2\nu-2\phi}
&=& y_0^2\,(\ddot{\beta}+\frac{2}{3}\dot{\beta}^2 -\dot{\beta}\dot{\nu}+\frac{1}{4}\,\dot{\phi}^2), \\
\label{MG0055ii}
\nu''-\frac{1}{3}\,\beta'^2+\frac{1}{4}\,\phi'^2-\a^2
e^{2\nu-2\phi}
&=& y_0^2 \,(\ddot{\nu}-\frac{1}{3}\,\dot{\beta}^2+\frac{1}{4}\,\dot{\phi}^2), \\
\label{MG05}
\dot{\beta}'+\frac{1}{3}\,\dot{\beta}\beta'-\dot{\nu}\beta'-\nu'\dot{\beta}+\half\,\dot{\phi}\phi'&=&
0. \eea

In the above, both the $G^5_5$ equation (\ref{MG55}) and the
$G_{05}$ equation (\ref{MG05}) involve only single derivatives
with respect to $\o$. Applying the junction conditions, we find
that both left-hand sides vanish when evaluated on the branes. The
$G_{05}$ equation is then trivially satisfied, while the $G^5_5$
equation yields the relation \be
\frac{b_{,TT}}{b}+\frac{b_{,T}^2}{b^2}-\frac{b_{,T}\,n_{,T}}{bn}+\frac{1}{12}\,\phi_{,T}^2 = 0, %\mid_{\o=\pm1}
\ee valid on both branes.  Introducing the brane conformal time
$\tau$, defined on either brane via $b \,y_0 \,\d \tau = n\,  \d
T$, this relation can be re-expressed as \be \label{Mbraneeq}
\frac{b_{,\tau\tau}}{b}+\frac{1}{12}\,\phi_{,\tau}^2=0. \ee

Notice that in the Einstein equations
(\ref{Mphi})-(\ref{MG0055ii}), all terms involving time
derivatives appear at $O(y_0^2)$ higher than the terms involving
derivatives along the orbifold direction. Thus, solving the
equations of motion perturbatively in powers of $y_0^2$, at zeroth
order we have a set of ordinary differential equations in $\w$.
Integrating these differential equations, we obtain the $\w$
dependence of the solution at leading order, along with three
arbitrary functions of time.  How these functions are determined
will be explained below in Section \ref{X}. Then, at
next-to-leading order, we have a set of ordinary differential
equations for the $\w$ dependence of the solution at $O(y_0^2)$,
involving source terms constructed from the time dependence at
leading order. Thus, solving the Einstein equations is reduced to
an iterative procedure involving the solution of a finite number
of ordinary differential equations at each new order in $y_0^2$.

\subsection{Initial conditions}

In order to fix the initial conditions for our expansion in
$y_0^2$, it is useful to solve for the bulk geometry about the
collision as a series expansion in $t$ (in this section we revert
briefly to our original $t$ and $y$ coordinates). Up to terms of
order $t^3$, the solution corresponding to the Kaluza-Klein zero
mode is: \bea
n &=& 1 + \a\,(\sech{y_0}\sinh{y})\,t + \frac{\a^2}{8}\,\sech^2 y_0\,(9-\cosh{2y_0}-8\cosh{2y})\,t^2, \qquad \\
b &=& 1 + \a\,(\sech{y_0}\sinh{y})\,t+\frac{\a^2}{4}\,\sech^2 y_0 (3+\cosh{2y_0}-4\cosh{2y})\, t^2, \qquad \\
e^\phi &=& 1+6\a\,(\sech{y_0}\sinh{y})\,t-\frac{3\a^2}{2}\,\sech^2
y_0 (2-\cosh{2y_0}-\cosh{2y})\, t^2, \qquad \eea where we have
used the junction conditions to fix the arbitrary constants
arising in the integration of the bulk equations with respect to
$y$. The brane conformal times $\tau_\pm$ (where the subscript
$\pm$ refers to the brane locations $\o = \pm 1$, or equivalently,
to their tensions) are then given by \be \tau_\pm = \int
\frac{n_\pm}{b_\pm}\,\d t =t + O(t^3), \ee in terms of which the
brane scale factors $b_\pm$ are \be \label{btau} b_\pm = 1 \pm
\a\,\tau_\pm \tanh{y_0}-\frac{3}{2}\,\a^2\tau_\pm^2 \tanh^2{y_0} +
O(\tau_\pm)^3. \ee

\subsection{A conserved quantity} \label{X}

As explained above, every time we integrate the bulk Einstein
equations at a given order in $y_0^2$ with respect to $\w$ we pick
up three arbitrary functions of time.  Two of these three
functions may be determined with the help of the $G^5_5$ equation
evaluated on both branes, namely (\ref{Mbraneeq}). To fix the
third arbitrary function, however, a further equation is needed,
which we derive below.

Introducing the variable $\chi=\phi-2\beta$, upon subtracting
twice (\ref{MG00p55}) from (\ref{Mphi}) we find \be
\label{chiwave} (\chi'e^\beta)' = y_0^2
(\dot{\chi}e^\beta)\dot{\,}, \ee which is the massless wave
equation $\bo\chi=0$ in this background.\footnote{For branes with
non-zero spatial curvature, however, there is an additional source
term in this equation, invalidating the argument that follows.}
Since the junction conditions imply $\chi'=0$ on the branes, the
left-hand side vanishes upon integrating over $\o$. A second
integration over $\ln T$ then yields \be \label{charge}
\int_{-1}^{+1} \d \o \,\dot{\chi}e^\beta = \gamma, \ee for some
constant $\gamma$, which we can set to zero since our initial
conditions are such that $\dot{\chi}e^\beta \tt 0$ as $T\tt 0$.

Let us now consider solving (\ref{chiwave}) for $\chi$, as a
perturbation expansion in $y_0^2$. Setting $\chi=\chi_0+y_0^2\,
\chi_1+O(y_0^4)$, and similarly for $\beta$, at zeroth order we
have $(\chi_0'e^{\beta_0})'=0$. Integrating with respect to $\o$
introduces an arbitrary function of $T$ which we can immediately
set to zero using the boundary condition on the branes, which,
when evaluated to this order, read $\chi_0'=0$.  This tells us
that $\chi_0'=0$ throughout the bulk; $\chi_0$ is then a function
of $T$ only, and can be taken outside the integral in
(\ref{charge}).  Since $\gamma=0$, yet the integral of $e^\beta$
across the bulk cannot vanish, it follows that $\chi_0$ must be a
constant.

At order $y_0^2$, the right-hand side of (\ref{chiwave}) evaluates
to $y_0^2 (\dot{\chi}_0e^{\beta_0})\dot{\,}$, which vanishes.
Evaluating the left-hand side, we have $(\chi_1'e^{\beta_0})'=0$,
and hence, by a sequence of steps analogous to those above, we
find that $\chi_1$ must also be constant.  It is easy to see that
this behavior continues to all orders in $y_0^2$.  We therefore
deduce that $\chi = \phi-2\beta$ is exactly constant. Since both
$\phi$ and $\beta$ tend to zero as $T\tt 0$, this constant must be
zero, and so we find
\[
\phi=2\beta.
\]

The essence of this result is that a perturbative solution in
powers of $y_0^2$ exists only when $\chi$ is in the Kaluza-Klein
zero mode. This may be seen from (\ref{chiwave}), which reduces,
in the limit where $T\tt 0$ and $\beta\tt 0$, to $\chi''=y_0^2
\ddot{\chi}$. The existence of a perturbative expansion in $y_0^2$
requires the right-hand side of this equation to vanish at leading
order. This is only the case, however, for the Kaluza-Klein zero
mode: all the higher modes have a rapid oscillatory time
dependence such that the right-hand side {\it does} contribute at
leading order. For these higher Kaluza-Klein modes, therefore, a
gradient expansion does not exist.

Setting $\phi=2\beta$ from now on, returning to (\ref{Mbraneeq})
and recalling that $\beta=3\ln b$, we immediately obtain the
equivalent of the Friedmann equation on the branes, namely \be
\Big(\frac{b_{,\tau}}{b}\Big)_{,\tau}+4\,\Big(\frac{b_{,\tau}}{b}\Big)^2=0.
\ee Integrating, the brane scale factors are given by \be b_\pm =
\bar{A}_\pm (\tau_\pm - c_\pm)^{1/4}. \ee To fix the arbitrary
constants $\bar{A}_\pm$ and $c_\pm$, we need only to expand the
above in powers of $\tau_\pm$ and compare with (\ref{btau}).  We
find \be \label{Mb} b_\pm = \big[1\pm
(4\a\tanh{y_0})\,\t_\pm\big]^{1/4}. \ee This equation determines
the brane scale factors to all orders in $y_0^2$ in terms of the
conformal time on each brane.  (As a straightforward check, it is
easy to confirm that the $O(\tau_\pm^2)$ terms in (\ref{btau}) are
correctly reproduced).

\subsection{The scaling solution}
\label{scalingsection}

We are now in a position to solve the bulk equations of motion
perturbatively in $y_0^2$. Setting \be \beta = \beta_0+y_0^2\,
\beta_1 + O(y_0^4), \qquad \nu = \nu_0+y_0^2\,\nu_1+O(y_0^4), \ee
the leading terms of this expansion (namely $\beta_0$ and $\nu_0$)
constitute a {\it scaling solution}, whose form is independent of
$y_0$ for all $y_0 \ll 1$. To determine the $\o$ dependence of
this scaling solution, we must solve the bulk equations of motion
at zeroth order in $y_0^2$. Evaluating the linear combination
$(\ref{MG00p55})-3[(\ref{MG55})+(\ref{MG0055ii})]$, noting that at
this order the right-hand sides all vanish automatically, we find
\be ((\beta'_0-3\nu'_0)e^{\beta_0})'=0. \ee Using the boundary
condition $\beta'_0-3\nu'_0=0$ on the branes, we then obtain
$\beta_0=3\nu_0+f(T)$, for some arbitrary function $f(T)$.
Substituting back into (\ref{MG55}) and taking the square root
then yields \be \nu'_0= \a\,e^{-5\nu_0-2f}, \ee where consistency
with the junction conditions (\ref{jct}) forced us to take the
positive root. Integrating a second time, and re-writing
$e^f=B^{-15}$, we find \be \label{wdep} e^{\nu_0} = B^6(T)\,
h^{1/5}, \qquad e^{\beta_0} = B^3(T)\, h^{3/5}, \ee where $h =
5\a\o+C(T)$, with $C(T)$ arbitrary.

In the special case where $B$ and $C$ are both constant, we
recover the exact static domain wall solution (\ref{domainwall}),
up to a trivial coordinate transformation. In general, however,
these two moduli will be time-dependent. Inverting the relation
$b_\pm^5 = B^5 (\pm5\a+C)$ to re-express $B$ and $C$ in terms of
the brane scale factors $b_\pm(T)$, we find \be \label{BCexp} B^5
= \frac{1}{10\a}\,(b_+^5 - b_-^5), \qquad C=5\a \left( \frac{b_+^5
+ b_-^5}{b_+^5 - b_-^5} \right) . \ee Furthermore, at zeroth order
in $y_0^2$, the conformal times on both branes are equal, since
\be y_0 \,\d \tau = \frac{n}{b}\,\d T =
e^{\nu_0-\beta_0/3}\,\frac{\d T}{T}=B^5(T)\,\frac{\d T}{T} \ee is
independent of $\o$. To this order then, (\ref{Mb}) reduces to \be
\label{bthing} b_\pm = (1\pm 4\a y_0 \tau)^{1/4}, \ee allowing us
to express the moduli in terms of $\tau$ as \bea \label{Bmod}
B^5(\tau) &=& \frac{1}{10\a}\,[(1+4\a y_0 \tau)^{5/4}-(1-4\a y_0 \tau)^{5/4}], \\ \nn \\
\label{Cmod} C(\tau) &=& 5\a\left[\frac{(1+4\a y_0
\tau)^{5/4}+(1-4\a y_0
    \tau)^{5/4}}{(1+4\a y_0 \tau)^{5/4}-(1-4\a y_0\tau)^{5/4}}\right].
\eea

The brane conformal time $\tau$ and the Milne time $T$ are then
related by \be \ln T = 10 \a y_0 \int [(1+4\a y_0
\tau)^{5/4}-(1-4\a y_0 \tau)^{5/4}]^{-1} \d\tau. \ee Rather than
attempting to evaluate this integral analytically, we will simply
adopt $\tau$ as our time coordinate\footnote{Note that the
relation between $T$ and $\tau$ is monotonic.  At small times $T =
y_0 \tau+(\a^2/4)y_0^3\tau^3+O(y_0^5\tau^5)$.} in place of $T$.
The complete scaling solution is then given by \bea
\d s^2 &=& h^{2/5}(\t,\o)\, \Big[B^2(\t)(-\d \t^2+ \d\vec{x}\,^2) + B^{12}(\t) \d\o^2\Big], \nn \\
e^{\f} &=& B^6(\t)\, h^{6/5}(\t,\o), \label{scalingsolution} \nn \\
h(\t,\o) &=& 5\a\w + C(\t), \eea with $B(\t)$ and $C(\t)$ given in
(\ref{Bmod}) and (\ref{Cmod}). This scaling solution solves the
bulk Einstein equations and the junction conditions to leading
order in our expansion in $y_0^2$.  The subleading corrections at
$y_0^2$ higher order may be obtained in an analogous fashion,
although we will not pursue them here.

The expressions above for the scaling solution can also be used to calculate the %resulting
%(time-dependent)
slope of the `warp factor' appearing in the metric
(\ref{scalingsolution}), namely $hB^5$. The slope, which from the
Israel matching condition represents the effective strength of the
two brane tensions, is given by \be (h
B^5)'=\frac{1}{2}\,\big[(1+4\a y_0\t)^{5/4}-(1-4\a
y_0\t)^{5/4}\big]. \label{slope} \ee Thus we can picture the brane
tensions to be evolving in time. In particular, note that what was
a downward-pointing kink before the collision turns into an
upward-pointing kink after the collision (and vice versa), while
at the collision itself the slope is zero. Thus the tension of the
branes swaps over at the collision, with both branes becoming
effectively tensionless at the collision itself.  (This is another
indication that the collision represents a rather mild
singularity).

The scaling solution is valid for times in the range $-(4\a
y_0)^{-1} < \tau < (4\a y_0)^{-1}$. At $\tau = \pm (4\a
y_0)^{-1}$, however, the scale factor on the negative-tension
brane vanishes from (\ref{bthing}) (recalling that the tension of
the branes is reversed for $\t < 0$). Since the physical
interpretation of these events is much clearer in the alternative
coordinate system in which the bulk is static and the branes are
moving, we will postpone a full discussion until Section
\ref{sectionBounce}. (In fact, it will turn out that to continue
the solution to times $\t > (4 \a y_0)^{-1}$, we simply need to
introduce absolute value signs around all factors of $(1-4\a y_0
\t)$).

A further quantity of interest, the distance between the branes,
evolves as \be d = \int_{-1}^{+1} h^{1/5} B^6 =
\frac{1}{6\a}\,[(1+4\a y_0\t)^{3/2}-(1-4\a y_0\t)^{3/2}]. \ee
Thus, for small $\t$ we have $d \simeq 2y_0 \t$, independent of
$\a$, and for large $\t$ (imposing an absolute value on the second
term), we have $d \simeq (y_0 \t/\a)^{1/2}$.

\subsection{Lifting to eleven dimensions}

It is straightforward to lift the scaling solution to eleven
dimensions, where the five-dimensional metric and scalar field are
both part of the eleven-dimensional metric. The 4-form field
strength has a non-zero component in the Calabi-Yau directions
only: \be G_{abcd} =
\frac{\a}{\sqrt{2}}\,{\e_{abcd}}^{ef}\,\o_{ef}, \qquad  y_0 < y <
y_0, \ee where $\o_{ef}$ is the K\"{a}hler form on the Calabi-Yau.
In the extended picture, the sign of $G_{abcd}$ reverses across
the branes. The metric in eleven dimensions reads \bea
\d s_{11}^2 &=& e^{-2\f/3}\d s_5^2 + e^{\f/3} \d s_{\mathrm{CY}}^2 \nn \\[1ex]
&=& h^{-2/5} B^{-2} \, ( -\d\t^2+\d\vec{x}\,^2 ) + h^{-2/5} B^8
\d\o^2 + h^{2/5} B^2 \d s_{\mathrm{CY}}^2, \eea and the 4-form
flux is proportional to $\a$ when all its indices are pointing in
a Calabi-Yau direction. Thus the eleven-dimensional distance
between the branes is given, to leading order in $y_0^2$, by \be
d_{11} = B^4 \int_{-1}^{+1} \d \o \,h^{-1/5} = 2y_0\t. \ee This
simple relationship underlies the utility of $\t$ as a time
coordinate.

Finally, the volume of the Calabi-Yau manifold, averaged over the
orbifold, is \be \langle e^{\f}\rangle = \frac{1}{2} B^6
\int_{-1}^{+1} \d \o\, h^{6/5} = \frac{5}{11} \left[ \frac{(1+4\a
y_0\t)^{11/4}-|1-4\a y_0\t|^{11/4}} {(1+4\a y_0\t)^{5/4}-|1-4\a
y_0\t|^{5/4}}\right], \ee where we have imposed the absolute
values to permit a continuation to late times, as will be
explained in Section \ref{sectionBounce}. Thus the radius of the
Calabi-Yau grows at large times as $\langle e^{\f/6}\rangle \sim
\t^{1/4}$, whereas the distance between the branes $d_{11}$ grows
linearly in $\t$.  A phenomenologically acceptable configuration
where $d_{11} \simeq 30\, e^{\f/6}$ \cite{BD} is therefore quite
naturally obtained, assuming that both the distance between the
branes and the Calabi-Yau volume modulus are stabilised by an
inter-brane potential when they reach large values. We will leave
a more detailed investigation of the eleven-dimensional properties
of our solution to future work.

\section{An alternative perspective:  moving branes in \\ a static bulk}

Let us now consider an alternative coordinate system in which the
bulk is static and the branes are moving. To find the bulk metric
in this coordinate system we can make use of a modified version of
Birkhoff's theorem, as shown in Appendix A. Assuming only
three-dimensional homogeneity and isotropy, as consistent with
cosmological symmetry on the branes, in addition to the exact
relation $\phi=2\beta$ (motivated in Section \ref{X} through
considerations of regularity at the brane collision), the bulk
Einstein equations can be integrated exactly.  One can then choose
the static parameterisation: \bea \label{staticbulk}
\d s^2 &=& -f\,\d t^2 + \frac{r^{12}}{f}\,d r^2 + r^2\, \d\vec{x}^2, \nn \\[1ex]
f(r) &=& \a^2 r^2 - \mu r^4, \qquad e^\phi = r^6, \eea where $0\le
r\le \sqrt{\mu}/\a$, and the coordinate $t$ is unrelated to the
Milne time appearing in the previous section. Physically, this
solution describes a timelike naked singularity located at $r=0$,
and was first discovered %(in different coordinates)
by Chamblin and Reall in \cite{Chamblin:1999ya} (who instead
looked directly for solutions in which the bulk was static). Since
the coordinates above do not cover the whole spacetime manifold,
the maximal extension may easily be constructed following the
usual Kruskal procedure, as detailed in Appendix C.

To find the trajectories of a pair of positive- and
negative-tension branes embedded in this static bulk spacetime we
solve the Israel matching conditions in the usual fashion. After
performing this calculation (see Appendix B, and also
\cite{Chamblin:1999ya}), one finds that the induced brane metrics
are indeed cosmological, with scale factors given by
\[
\label{bscalefactors} b_\pm = (1\pm 4\sqrt{\mu}\, \t_\pm)^{1/4},
\]
where $\tau_\pm$ is the brane conformal time, and we have rescaled
$b_\pm$ to unity at the collision, which is taken to occur at
$\tau_\pm = 0$.

Upon setting $\mu=0$, we immediately recover the static domain
wall solution (\ref{domainwall}), after a suitable change of
coordinates. More generally, we require $\mu\ge 0$ to avoid the
appearance of imaginary scale factors. Through comparison with our
earlier result (\ref{Mb}), we also find
\[
\label{lambdadef} \sqrt{\mu} = \a\,\tanh{y_0}.
\]
Thus, for $\mu >0$, after starting off coincident, the two branes
proceed to separate.  However, while the positive-tension brane
travels out to large radii unchecked, the negative-tension brane
reaches the naked singularity (at which $b_-$ and $e^\phi$ tend to
zero) in a finite brane conformal time $\tau_- =
(4\sqrt{\mu})^{-1}$.

In the following section we will argue that the resulting
singularity is extremely mild, and simple to regularise. If almost
any type of well-behaved matter is present on the negative-tension
brane -- even in only vanishing quantities -- then, rather than
hitting the singularity, the negative-tension brane will instead
undergo a bounce at some small finite value of the scale factor
and move away from the singularity.

\subsection{The bounce of the negative-tension brane} \label{sectionBounce}

The Friedmann equation for the negative-tension brane is derived %%V2
in Appendix B. For the case in which a time-dependent scalar field (with an
$e^{\phi}$ coupling to the Calabi-Yau volume scalar) is present on the brane, this
equation takes the form (see (\ref{F2}))
\[
H_-^2 = \frac{\dot{b}_-^2}{b_-^2}= -
2 \a \,\frac{A_-}{b_-^{18}}+ \frac{A_-^2}{b_-^{24}}+\frac{\mu}{b_-^{10}},
\]
where the constant $A_-$ parameterises the scalar kinetic energy
density.

The key feature of this equation is the negative sign in front of
the first term on the right-hand side, reflecting the fact that
matter on the negative-tension brane couples to gravity with the
wrong sign.\footnote{This property is a general feature of
braneworld models, see \eg \cite{Shiromizu}.}
For sufficiently large values of the scale factor the right-hand
side is dominated by the $\mu\, b_-^{-10}$ term. If we further
assume that the matter density on the branes is small compared to
the brane tension\footnote{As is in any case necessary for the
existence of a four-dimensional effective description.}  (\ie $A_-
\ll \a$),
so that the term linear in $A_-$ dominates over the quadratic term, then %(by virtue of the negative sign of this linear term)
it follows that at some small value of the scale factor the entire
right-hand side must vanish. Thus a negative-tension brane,
initially travelling towards the singularity, will generically
undergo a smooth bounce at some small value of the scale factor.
After this bounce the brane travels away from the singularity back
towards large values of the scale factor. This behaviour is
specific to the negative-tension brane, and moreover, persists
even in the limit in which $A_-$ (and hence the initial matter
density) is negligibly small.\footnote{In the simplest version of
  heterotic M-theory, the scalars present on the negative-tension
  brane do not couple to $\phi$ \cite{LOSW2}. Nevertheless, in the limit of small matter density on the brane, it can be shown that the
  brane bounces in a manner identical to the case described
  above. When the scalars do not couple to $\phi$, there are corrections to the bulk
  geometry in the vicinity of the bounce, but these become negligibly small as the scalar field density decreases to
  zero.}%%V2

In fact, even in the complete absence of matter on the
negative-tension brane, the bounce off the naked singularity is
still a relatively smooth event: converting to Kruskal coordinates
so that light rays are at $\pm \pi/4$ angles, using our exact
solution we show in Appendix C that the trajectory of the
negative-tension brane becomes precisely tangential to the
singularity at the moment of the bounce. In these coordinates
then, the negative-tension brane merely grazes the singularity
with zero velocity, before moving away again according to a
well-defined smooth continuation. The complete solution is
illustrated in Figures \ref{Kruskal} and \ref{doubleK}.

\begin{figure}[t]
\begin{center}
\psfrag{b4}{$b^4$} \psfrag{t}{$\a y_0\t$}
\includegraphics[width=12cm]{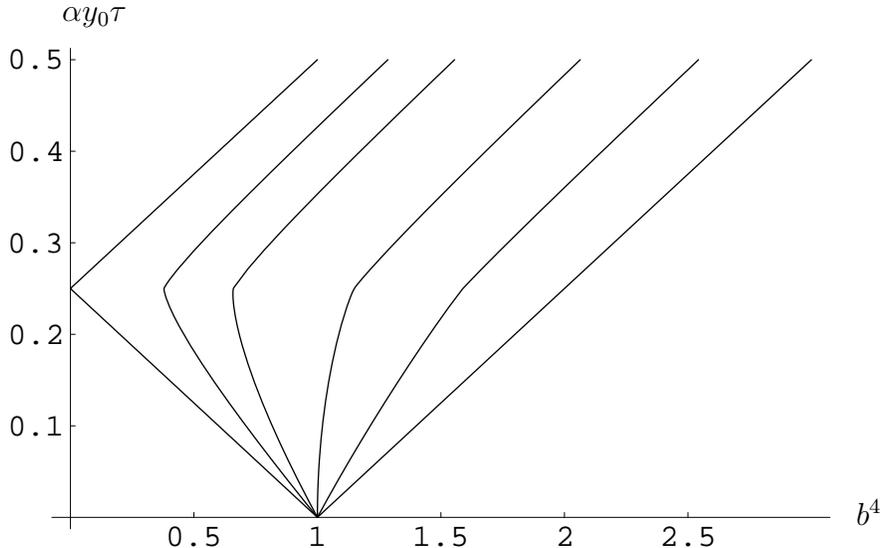}
\caption{ \label{coathanger} {\small Continuation of the scaling
solution (accurate to leading order in $y_0$) beyond the bounce of
the negative-tension brane at $\a y_0\t=1/4$, using the absolute
value prescription. Contours of constant $\w$ are plotted against
the fourth power of the three-dimensional scale factor $b$ and the
brane conformal time $\t$. The contours shown are, from left to
right, $\w = -1$, $-0.75$, $-0.5$, $0$, $0.5$, and $1$.  Thus the
left-most and right-most trajectories correspond to the negative-
and positive-tension branes respectively (for which $b_\pm^4 =
|1\pm 4 \a y_0 \t|$), with the region in between representing the
bulk. } }
\end{center}
\end{figure}

In this solution, the scale factor on the negative-tension brane
is given exactly for all times post-collision by
\[
b_- = |1-4 \sqrt{\mu} \,\t_-|^{1/4},
\]
where $\t_-$ is the conformal time on the negative-tension brane.
This result shows us how to continue our earlier scaling solution
in brane-comoving coordinates (see Section \ref{scalingsection})
to times after the bounce of the negative-tension brane: we simply
insert an absolute value sign into the expression (\ref{bthing})
for $b_-$ (recalling that $\tau$ is indeed the brane conformal
time in these coordinates).  The moduli $B(\tau)$ and $C(\tau)$
then follow from (\ref{BCexp}), which amounts to replacing all
factors of $(1-4\a y_0 \t)$ in (\ref{Bmod}) and (\ref{Cmod}) with
$|1-4 \a y_0 \t|$. It is easy to check that this continuation does
not affect the smooth evolution of quantities defined on the
positive-tension brane, and that the range of the coordinate $\w$
along the extra dimension is unaltered by this continuation (see
Figure \ref{coathanger}). Nevertheless, this brane-comoving
coordinate system is not particularly well-adapted to deal with
the actual moment of the bounce itself, as evidenced by the
`kinks' in Figure \ref{coathanger}. With the presence of
regulatory matter on the negative-tension brane, however, the
bounce will occur at some small non-zero $b_-$, smoothing out
these kinks.

To check that the scaling solution is qualitatively correct near
the bounce of the negative-tension brane, it is interesting to
examine the leading behaviour of the exact bulk metric about the
singularity. Setting $r^5 = 5\rho + c$, for some constant $c$, the
metric (\ref{staticbulk}) becomes \be \d s^2 =
(5\rho+c)^{2/5}\left[-\big(\a^2-\mu (5\rho +c)^{2/5}\big)\d t^2 +
  \d\vec{x}^2 + \frac{\d \rho^2}{\a^2-\mu (5\rho+c)^{2/5}}\right],
\ee which, after trivial re-scalings, to leading order in $r$
reads \be \d s^2 =(5\rho+c)^{2/5}[-\d t^2 + \d\vec{x}^2 + \d
\rho^2]. \ee This is precisely the form of our scaling solution
(\ref{scalingsolution}) at a given instant in time.

Note also that in the exact solution, the proper velocity of the
negative-tension brane vanishes as it approaches the naked
singularity: from (\ref{staticbulk}) the proper velocity is $r^6
\d r/ (f \d t)$, which scales as $r^4 \d r/\d t$ at small $r$. Yet
from (\ref{rteom}), $\d r/\d t$ scales as $f/r^5 \sim 1/r^3$, and
so the proper velocity tends to zero linearly with $r$. This
result, that the proper velocity of the negative-tension brane
tends to zero linearly with $r$ as it approaches the naked
singularity, is also found in Kruskal coordinates, as shown in
(\ref{dVdU}).

\section{Conclusions}

We have presented a cosmological solution describing the collision
of the two flat boundary branes in heterotic M-theory. This
solution is a significant step towards our goal of describing the
cosmic singularity as a brane collision within the well-motivated
framework of Ho\v{r}ava-Witten theory. Requiring the collision to
be the `least singular' possible, {\it i.e.}, that the metric
tends towards (2d compactified Milne)$/\mathbb{Z}_2 \times
\mathbb{R}_3$ times a finite-volume Calabi-Yau, has two important
consequences. First, it selects a single solution to the equations
of motion. Second, it shifts the singularity in the Calabi-Yau
volume that one might have naively expected at the brane collision
to two spacetime events before and after the brane collision. We
have shown these two events to be very mild singularities, which
are easily removed by including an arbitrarily small amount of
matter (for example scalar field kinetic energy) on the %%V2
negative-tension brane. Before the initial bounce of the
negative-tension brane, and after the final bounce, the solution
presented here can be identified with that described by Chen {\it
et al.~}\cite{Gary}.

When the branes move at a small velocity, we expect to be able to
accurately describe the solution using a four-dimensional
effective theory (see \eg \cite{KSHW, GaugesII, Gonzalo1, Gonzalo2, Koyama}
and also \cite{McFadden}). We shall present such a description in
a companion publication \cite{LT}.

If our colliding brane solution is to successfully describe the
universe, we must also add potentials capable of stabilising the
moduli; in particular the volume of the Calabi-Yau manifold, which
determines the value of gauge couplings, and the distance between
the branes, which determines Newton's constant of gravitation.
These potentials also permit us to generate an interesting
spectrum of cosmological perturbations. Although the required
potentials cannot yet be derived from first principles, we can
study the consequences of various simple assumed forms. The
results will be presented elsewhere \cite{LMTS}.

\begin{center}
***
\end{center}

{\small {\it Acknowledgements:}  The authors wish to thank Mariusz
Dabrowski, Gary Gibbons, Andr{\'e} Lukas, Ian Moss, Chris Pope, Paul
Steinhardt, Kelly Stelle and Andrew Tolley for useful discussions.
JLL and NT acknowledge the support of PPARC and of the Centre for
Theoretical Cosmology, in Cambridge. PLM is supported through a
%Foundation for Fundamental Research on Matter (FOM) and the
Spinoza Grant of the Dutch Science Organisation (NWO). }

\appendix
 %{\bf Appendix}
\renewcommand{\theequation}{\Alph{section}.\arabic{equation}}
\setcounter{equation}{0}

\section{A modified Birkhoff theorem}

In this appendix we show how the assumption of cosmological
symmetry on the branes, coupled with the exact result
$\phi=2\beta$ (derived in Section \ref{X} following from our
assumption that the Riemann curvature remains bounded as we
approach the collision), is sufficient to uniquely determine the
bulk metric.  Moreover, with an appropriate choice of
time-slicing, the bulk geometry can be written in static form. Our
derivation closely parallels the analogous case for the
Randall-Sundrum model \cite{Gregory}.

We start with the fully-general metric ansatz
\[
\d s^2 = e^{2\sigma-2\psi/3}(-\d t^2 + \d y^2)+e^{2\psi/3}\d
\Omega_3^2,
\]
where $\sigma(t,y)$ and $\psi(t,y)$ are arbitrary functions, and
the coordinates $t$ and $y$ are fresh, unrelated to those
appearing earlier.  To allow for the possibility of open, flat or
closed brane geometries, we will take
\[
\d \Omega_3^2 = (1-k\rho^2)^{-1}\d \rho^2 + \rho^2 \d\Omega_2^2,
\]
where $k=-1$, $0$ or $+1$ respectively, and $\d\Omega_2^2$ denotes
the usual line element on a unit two-sphere. The above choice
represents the most general bulk metric consistent with
three-dimensional spatial homogeneity and isotropy, after we have
made use of our freedom to write the ($t$, $y$) part of the metric
in a conformally flat form.

In light of Section \ref{X}, we will additionally set $\phi = 2
\psi$. Then, evaluating the Einstein equations
(\ref{Mphi})-(\ref{MG05}) in the bulk, after taking appropriate
linear combinations we find \bea
\psi_{,ty} &=& \sigma_{,y}\psi_{,t}+\sigma_{,t}\psi_{,y}-3\psi_{,t}\psi_{,y}, \\
\psi_{,tt}+\psi_{,yy} &=& 2 \sigma_{,t}\psi_{,t}+2\sigma_{,y}\psi_{,y}-3\psi_{,t}^2-3\psi_{,y}^2, \\
\psi_{,tt}-\psi_{,yy} &=& - \psi_{,t}^2+\psi_{,y}^2+6 \a^2 e^{2\sigma-14 \psi/3}, \\
\sigma_{,tt}-\sigma_{,yy} &=& -\psi_{,t}^2+\psi_{,y}^2+\a^2
e^{2\sigma-14\psi/3}+2 k e^{2\sigma-4\psi/3}. \eea Switching to
the light-cone coordinates,
\[
u = \frac{1}{2}\,(t-y), \qquad v =\frac{1}{2}\,(t+y),
\]
these become \bea
\psi_{,uu} &=& 2\sigma_{,u}\psi_{,u}-3 \psi_{,u}^2, \\
\psi_{,vv} &=& 2\sigma_{,v}\psi_{,v}-3 \psi_{,v}^2, \\
\psi_{,uv} &=& -\psi_{,u}\psi_{,v}+6 \a^2 e^{2\sigma-14\psi/3}, \label{Beq3} \\
\sigma_{,uv} &=& -\psi_{,u}\psi_{,v}+\a^2 e^{2\sigma-14\psi/3}+2k
e^{2\sigma-4\psi/3}. \label{Beq4} \eea Integrating the first two
equations, we find
\[
e^{2\sigma-3\psi}=\psi_{,u}V'(v)=\psi_{,v} U'(u),
\]
where $U'(u)$ and $V'(v)$ are arbitrary non-zero functions, and we
will use primes to denote ordinary differentiation wherever the
argument of a function is unique.  It follows that $\psi$ and
$\sigma$ take the form
\[
\psi=\psi(U(u)+V(v)),   \qquad e^{2\sigma-3\psi}=\psi' U' V'.
\]
Consistency between (\ref{Beq3}) and (\ref{Beq4}) then requires
$k=0$, and so only spatially flat three-geometries are
permitted.\footnote{Similar
  conclusions were reached by Chamblin and Reall in \cite{Chamblin:1999ya}.}
Integrating once, we find
\[
\psi' e^\psi = 9\mu - 9\a^2 e^{-2\psi/3},
\]
where $\mu$ is an arbitrary constant.

The metric
\[
\d s^2 = -4 \psi' e^{7\psi/3}\d U \d V+e^{2\psi/3}\d \vec{x}^2,
\]
upon changing coordinates to
\[
r=e^{\psi/3}, \qquad t=3(V-U),
\]
then takes the static form
\[
\label{staticbulk2} \d s^2 = -(\a^2 r^2 - \mu r^4)\,\d t^2 +
\frac{r^{12} \d r^2}{(\a^2 r^2-\mu r^4)} + r^2 \d\vec{x}^2.
\]
The corresponding Calabi-Yau volume is given by
\[
e^\phi = r^6.
\]

\section{Brane trajectories}

In this appendix we embed a pair of moving branes into the static
bulk spacetime (\ref{staticbulk2}) and derive the
Friedmann equations describing their trajectories. This is
accomplished by solving the Israel matching conditions
\cite{Israel}:
\[
\label{IMC} K^\pm_{ab} =
-\frac{1}{2}\,\big(T^\pm_{ab}-\frac{1}{3}\,g^\pm_{ab}\,
T^\pm\big),
\]
where $K^\pm_{ab}$, $T^\pm_{ab}$ and $g^\pm_{ab}$ denote the brane
extrinsic curvature, stress tensor and induced metric
respectively, and we are assuming a $\mathbb{Z}_2$ symmetry about
each brane.

Parameterising the trajectory of a given brane as $t=T\,(t_p)$ and
$r=R\,(t_p)$, where $t_p$ is the brane proper time, the induced
metric is
\[
\d s^2 = -\d t_p^2 + R^2(t_p)\, \d \vec{x}^2,
\]
hence $R$ may be associated with the relevant brane scale factor
$b_\pm$. The brane 4-velocity is then $u^a
=(\dot{T},\,\dot{R},\,\vec{0})$, where, throughout this appendix,
we will use dots to indicate differentiation with respect to
$t_p$.   The constraint $u^a u_a=-1$ yields the additional
relation $\dot{T} = (f+R^{12} \dot{R}^2)^{1/2}/f$. Similarly, the
unit normal vector $n^a$ is given by
\[
\label{normal} n_a = \pm (R^6 \dot{R},\, -\frac{R^6}{f}\,(f+R^{12}
\dot{R}^2)^{1/2}, \,\vec{0}),
\]
where the function $f(r)$ is as defined in (\ref{staticbulk}).  (Note that the choice of sign %in (\ref{normal})
corresponds to our choice of which side of the bulk we keep prior
to imposing the $\mathbb{Z}_2$ symmetry. Keeping the side for
which $r \le R(t_p)$ leads to the creation of a positive-tension
brane, with normal pointing in the direction of decreasing $r$,
requiring the positive sign for $n_a$. Conversely, if we retain
the $r\ge R(t_p)$ side of the bulk creating a negative-tension
brane, the normal points in the direction of increasing $r$ and we
must take the negative sign for $n_a$).

The three-spatial components of the brane extrinsic curvature are
then
\[
K^\pm_{ij} = \nabla_i n_j^\pm = \mp R^{-7}(f+R^{12}
\dot{R}^2)^{1/2}\, g^\pm_{ij},
\]
while the brane stress-energy is given by $T^\pm_{ab}=\mp 6\a
e^{-\phi} g^\pm_{ab}$. The Israel matching condition then yields
the Friedmann equation $H_\pm^2 =
\dot{R}^2/R^2=\dot{b}_\pm^2/b_\pm^2 =\mu/b_\pm^{10}$. Integrating,
we find \cite{Chamblin:1999ya}
\[
b_\pm = (1\pm 5\sqrt{\mu}\, t_p^\pm)^{1/5},
\]
where the constants of integration have been fixed by re-scaling
$b_\pm$ to unity at the collision, which is taken to occur at
$t_p^\pm=0$. In terms of brane conformal time $\tau_\pm$, this
relation reads
\[
b_\pm = (1\pm 4\sqrt{\mu} \,\tau_\pm)^{1/4},
\]
where the origins of $\t_\pm$ are chosen so that the branes
collide at $\t_\pm=0$.

This result may easily be generalised to scenarios in which matter %%V2
is incorporated on one or both of the branes. For example, in the
case where we add a scalar field $\chi$ on each brane, allowing for an
arbitrary coupling $F(\phi)$ to the Calabi-Yau volume modulus $\phi$, the
action should be augmented by the terms \be - \int_{4d, \, \omega =\pm 1}
\sqrt{-g^\pm}\, \frac{1}{2} \,(\partial \chi)^2 \,F(\phi) . \ee The brane
stress-energy is now
\[ T^\pm_{ab} = \mp 6 \a e^{-\phi}g^\pm_{ab}+
\frac{1}{2} F(\phi)\, [\chi_{,a} \chi_{,b} - \frac{1}{2}\,g^\pm_{ab}\,
(\partial \chi)^2].
\]
Since we are interested in cosmological solutions, we will
consider the scalar field $\chi$ to be a function of time only.
Evaluating the three-spatial components of (\ref{IMC}) then yields
the modified Friedmann equation
\[
\label{F1}
H_\pm^2 %= \frac{\dot{b}_\pm^2}{b_\pm^2}
= \pm \frac{\a}{12}\,\frac{\dot{\chi}^2 F}{b_\pm^6}
+\frac{\dot{\chi}^4 F^2}{24^2} +\frac{\mu}{b_\pm^{10}}.
\]
To find the dependence of $\chi$ on the scale factor $b_\pm$, it
is necessary to evaluate the $t_p t_p$ component of (\ref{IMC}).
This yields the equivalent of the usual cosmological energy
conservation equations. We start with
\[
K_{t_p t_p} = K_{ab}u^a u^b = u^a u^b \nabla_a n_b = -n_c a^c,
\]
where the acceleration $a^c = u^b \nabla_b u^c$.  Since $a^c u_c =
0$, the acceleration may also be written as $a^c = a n^c$, where
$a = -K_{t_p t_p}$. Then, since $\pt_t$ is a Killing vector of the
background\footnote{For a Killing vector $\xi^c$, $a n_\xi = a n_c
\xi^c = a_c \xi^c= \xi^c u^b \nabla_b u_c = u^b \nabla_b (u_c
\xi^c)-u^c u^b \nabla_b \xi_c$, where the last term vanishes by
Killing's equation, $\nabla_{(b} \xi_{c)}=0$, hence $a n_\xi = u^b
\pt_b u_\xi = \dot{u}_\xi$.}, we have $a = \dot{u}_t/n_t$. Thus
(\ref{IMC}) reads
\[
K^\pm_{t_p t_p} = \pm \frac{1}{R^6 \dot{R}}\, \frac{\d}{\d t_p} \,
\big( \a R \pm \frac{1}{24}\, R^7 \dot{\chi}^2 F)
= \pm \frac{\a}{R^6} -\frac{5}{24}\,\dot{\chi}^2 F,  %%V2
\]
where we have used (\ref{F1}). This leads to
\[
R \,(\dot{\chi}^2 F)_{,R} = -12 \,\dot{\chi}^2 F,
\]
{\it i.e.~}we have $\dot{\chi}^2 F = 24\, A_\pm\, R^{-12}$, for some
constant $A_\pm$. Thus the Friedmann equation becomes
\[ \label{F2}
H_\pm^2 %= \frac{\dot{b}_\pm^2}{b_\pm^2}
= \pm \frac{2 \a A_\pm}{b_\pm^{18}} +\frac{A_\pm^2}{b_\pm^{24}}
+\frac{\mu}{b_\pm^{10}}.
\]
We must also satisfy the junction condition arising from the
$\phi$ equation of motion. This is given by
\[
2\, n^a \partial_a \phi_{\pm} = \mp 12 \alpha \,e^{-\phi}_{\pm} -
\frac{1}{2}\, \dot{\chi}^2 F_{,\phi},
\]
which leads to
\[
H_\pm^2 %= \frac{\dot{b}_\pm^2}{b_\pm^2}
= \pm \frac{\a}{12}\,\frac{\dot{\chi}^2 F_{,\phi}}{b_\pm^6}
+\frac{\dot{\chi}^4 F_{,\phi}^2}{24^2} +\frac{\mu}{b_\pm^{10}}.
\]
Consistency with (\ref{F1}) then requires the coupling to be 
\[
F(\phi) = e^{\phi}.
\]
It can be verified that this coupling is also consistent with the
equation of motion for $\chi$.

Another form of brane-bound matter that alters the Friedmann %%V2
equations on the branes, while remaining consistent with the bulk
geometry, is a cosmological constant
$\L$ with $e^{-\phi}$ coupling to the Calabi-Yau volume scalar. In
this case, the modified Friedmann equation is given by
\[
H_\pm^2 = \pm \frac{\a}{3}\,\frac{\L}{b_\pm^{12}}+\frac{1}{36}\,\frac{\L^2}{b_\pm^{12}}+\frac{\mu}{b_-^{10}}.
\]

We will use the modified Friedmann equations (which are generally
similar in form to (\ref{F2})) in
Section \ref{sectionBounce}, to understand the bounce of the
negative-tension brane in the vicinity of the singularity.

\section{Kruskal extension of the bulk geometry}

\begin{figure}[t]
\begin{center}
\psfrag{UV}{$UV$} \psfrag{r}{$\rb$}
\includegraphics[width=10cm]{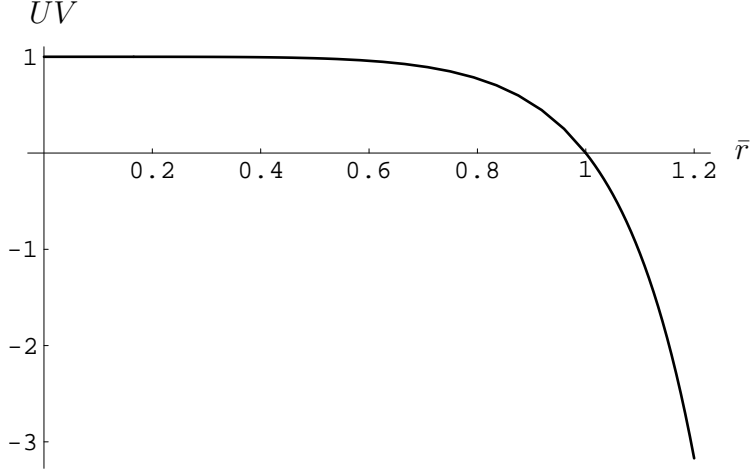}
\caption{ \label{UVr} {\small $UV$ plotted as a function of $\rb$,
according to (\ref{UVrel}) (taking $\lambda=1$). Note in
particular the smooth continuation to $\rb>1$ generated by the
sign flip in $\sigma$. The surfaces of constant $\rb$ correspond
to the hyperbolae $UV=\mathrm{constant}$, with this constant being
positive for $0\le \rb < 1$, and negative for $\rb >1$. } }
\end{center}
\end{figure}

In this appendix we consider the maximal extension of the bulk
geometry (\ref{staticbulk}), beyond the range $0\le r \le
\sqrt{\mu}/\a$ for which $f\ge 0$. These calculations will lead us
to the Kruskal diagram in Figure \ref{Kruskal}. Following Chen
{\it et al.}~\cite{Gary}, we start with the Eddington-Finkelstein
coordinates $u$ and $v$, defined via
\[
\label{eddfink} \d u = 4\a^2 \big(\d t + \frac{r^6}{f}\,\d r\big),
\qquad \d v = 4\a^2 \big(\d t - \frac{r^6}{f}\,\d r\big).
\]
The entire spacetime manifold is then covered by the Kruskal
coordinates
\[
U = \exp\( -\frac{\lambda^2}{4}\, u \), \qquad V = \exp
\(\frac{\lambda^2}{4}\, v\),
\]
where, for later convenience, we have introduced the constant
$\lambda = \sqrt{\mu}/\a = \tanh y_0$. In these coordinates, the
metric then reads \be \label{newmetric} \d s^2 = \frac{f}{\a^4
\lambda^4}\,\frac{\d U \d V}{UV} +r^2 \d\vec{x}^2. \ee Note that
$r$ should be understood here as a function of $UV$, given
implicitly by
\[
\label{UVrel} UV = \sigma \exp \left[
\frac{2}{\lambda^3}\,\left(\rb+\frac{1}{3}\,\rb^3-\frac{1}{2}\,\ln\left|\frac{1+\rb}{1-\rb}\right|\right)\right],
\]
where the re-scaled radial coordinate $\rb = \lambda r$, and
$\sigma$ is defined such that $\sigma=+1$ for $0\le \rb \le 1$,
while $\sigma=-1$ for $\rb > 1$. (This choice generates the smooth
continuation illustrated in Figure \ref{UVr}).
Surfaces of constant $\rb$ thus correspond to the hyperbolae $UV =
\mathrm{constant}$. In particular, we have chosen the constant of
integration deriving from (\ref{eddfink}) such that the
singularity at $\rb=0$ maps to the hyperbola $UV = 1$. The second
constant of integration implicit in (\ref{eddfink}) may then be
fixed by requiring that the time slice $t=0$ corresponds to the
line $U=V$ for $0\le \rb \le 1$, in which case
$V/U=\exp(2\lambda^2 \a^2 t)$.

From the results of the preceding appendix, the trajectory of a
brane (in the absence of any additional matter) is given, for
$0\le \rb \le 1$, by \be \label{rteom} \frac{\d r}{\d t} =
\pm\frac{\lambda f}{r^5}, \ee with the upper and lower signs
corresponding respectively to a brane moving away from, and a
brane moving towards, the singularity. Specialising to the case of
two branes moving in opposite directions, upon integrating we find
\[
\label{rttraj} \mp 2 \lambda^5 \a^2 t = \rb^2+\ln (1-\rb^2)
-\lambda^2-\ln (1-\lambda^2),
\]
where we have fixed the constants of integration by requiring that
the brane scale factors are unity at the collision (hence
$\rb=\lambda$ at $t=0$).

Converting now to Kruskal coordinates, we find the trajectories of
the branes are given parametrically, for all $0\le \rb< \infty$,
by \bea \label{bb1}
U &=& \cF(\rb),\qquad V=\cG(\rb) \qquad  \mathrm{(outgoing)}; \\
\label{bb2} U &=& \cG(\rb),\qquad V=\cF(\rb) \qquad
\mathrm{(infalling)}; \eea where the smooth functions \bea
\label{cFdef} \cF(\rb) &=& \sigma \exp \Big[
\frac{1}{\lambda^3}\,\big(\rb+\frac{1}{2}\,\rb^2+\frac{1}{3}\,\rb^3
+\ln |1-\rb|-\frac{1}{2}\,\lambda^2-\frac{1}{2}\,\ln(1-\lambda^2)\big)\Big], \\[1ex]
\cG (\rb) &=&\ \  \exp \Big[
\frac{1}{\lambda^3}\,\big(\rb-\frac{1}{2}\,\rb^2+\frac{1}{3}\,\rb^3
-\ln
(1+\rb)+\frac{1}{2}\,\lambda^2+\frac{1}{2}\,\ln(1-\lambda^2)\big)\Big].
\eea As previously, $\sigma=+1$ for $0\le \rb \le 1$, but
$\sigma=-1$ for $\rb > 1$, ensuring that $UV < 0$ for the portion
of the trajectories parameterised by $\rb > 1$.

\begin{figure}
\begin{center}
\includegraphics[width=11cm]{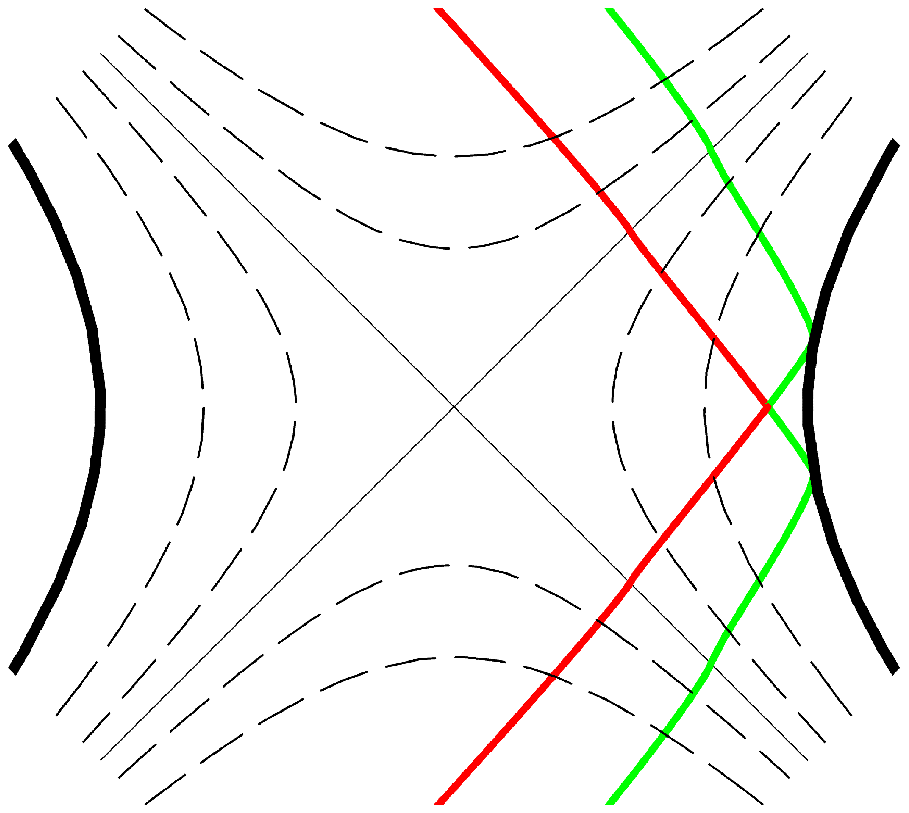} \\
\vspace{1cm}
\includegraphics[width=11cm]{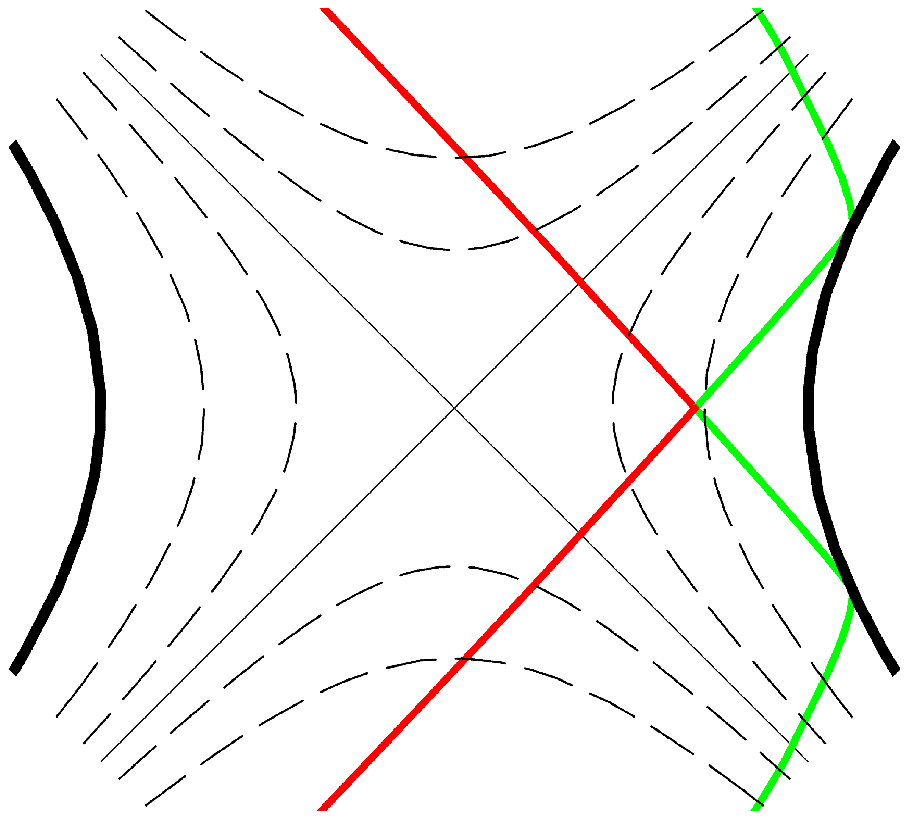}
\end{center}
\caption{ \label{doubleK} {\small Kruskal diagrams illustrating
the brane trajectories for
  different collision rapidities.  In the upper plot $y_0=1$,
and in the lower plot $y_0=1.5$. The corresponding plot for
$y_0=0.5$ appeared previously in Figure \ref{Kruskal}.} }
\end{figure}

To extend the brane trajectories beyond the bounce off the naked
singularity, we first of all write down the `extended'
trajectories in terms of the original $\rb$ and $t$ coordinates.
Altering the constants of integration in (\ref{rttraj}) so that
the different branches of the trajectories match up at the bounce,
we find
\[
\mp 2 \lambda^5 \a^2 t =\rb^2 +\ln (1-\rb^2) +\lambda^2+\ln
(1-\lambda^2),
\]
for $0\le\rb\le 1$, where the upper and lower signs correspond
respectively to an outgoing, and an infalling, brane. Converting
to Kruskal coordinates, this becomes \bea \label{ab1}
U &=& \cF(\rb)\,\cG^2(0),\qquad V=\cG(\rb)\,\cF^2(0) \qquad  \mathrm{(outgoing)}; \\
\label{ab2} U &=& \cG(\rb)\,\cF^2(0),\qquad V=\cF(\rb)\,\cG^2(0)
\qquad  \mathrm{(infalling)}; \eea where $\rb$ now takes values in
the entire range $0\le r < \infty$. Finally, plotting (\ref{bb1}),
(\ref{bb2}), (\ref{ab1}) and (\ref{ab2}), we obtain the full
Kruskal diagrams shown in Figure \ref{Kruskal} (for $y_0=0.5$),
and in Figure \ref{doubleK} (for $y_0=1$ and $1.5$).
An interesting feature of the brane trajectories revealed by these
plots are the points of inflection that occur whenever the brane
trajectories intersect the Boyer axes $UV=0$ (see Figure
\ref{Kruskal} especially). These may be understood as a
consequence of the vanishing of $f$ in (\ref{rteom}) whenever
$\rb=1$.

Note also that at the singularity ($\rb=0$ corresponding to
$UV=1$), the slope of the negative-tension brane trajectory is
exactly $-1/U^2$.  For example, using (\ref{bb1}), \be
\label{dVdU} \frac{\d V}{\d U} =
-\(\frac{1-\rb}{1+\rb}\)\frac{V}{U}, \ee which reduces to $-1/U^2$
when $\rb=0$ and $UV=1$.  A similar conclusion applies for each of
the other solution branches (\ref{bb2}), (\ref{ab1}), (\ref{ab2}).
Consequently, the brane trajectory at the bounce is tangent to the
singularity itself.  This means that the brane simply grazes the
singularity with vanishing normal velocity. Similarly for the
Calabi-Yau volume at the branes, $V_{\mathrm{CY}} = r^6$, we find
\[
\frac{\d V_{\mathrm{CY}}}{\d t} = \pm 6 \lambda f = \pm
6\lambda\a^2 r^2+O(r^4),
\]
and hence the rate of change of the Calabi-Yau volume on the
negative-tension brane, as it bounces off the singularity, is
zero. The above considerations suggest that the bounce of the
negative-tension brane is a relatively smooth event, even in the
absence of regulatory matter on the brane.

\newpage

%%%%%%%%%%%%%% END %%%%%%%%%%%%%%%%%%%%%%%%%%%%

\bibliographystyle{apsrev}
\bibliography{HWdraft12}

\end{document}